\documentclass[article,  onecolumn,showpacs,preprintnumbers,amsmath,amssymb]{revtex4}

\usepackage{graphicx}% Include figure files
\usepackage{dcolumn}% Align table columns on decimal point
\usepackage{bm}% bold math

\begin{document}
\author{Xiaohua Wu}
\address{Department of Physics, Sichuan University, Chengdu 610064, China.}
\title{ The Vector Representation of   Standard Quantum Process Tomography }
\begin{abstract}
    The characterization of the evolution
of a quantum system is one of the main tasks to accomplish to
achieve quantum information processing. The  standard quantum process tomography (SQPT) has the unique property that it can be applied without introducing any additional quantum  resources. In present work, we shall focus on the following two topics about the SQPT. At first, in the SQPT protocol for   a $d$-dimensional system, one should encounter a problem in solving  of a set of $d^4$ linear equations in order to get the matrix containing the complete information about the unknown quantum channel.  Until now, the general form of the solution is unknown. And a long existed conviction is that the solutions are not unique. Here, we shall develop a self-consistent scheme,  in which bounded linear operators are presented by vectors, to construct the set of linear equations. With the famous Cramer's rule for the  set of linear equations, we are able to  give the general form of the solution and prove that it is unique.
In second, the central idea of the SQPT is to prepare a set of linearly independent inputs and measuring the outputs via the quantum state tomography (QST).   Letting  the inputs and the measurements be prepared by  two sets of the rank-one
   positive-operator-valued measures [POVMs], where each POVM  is supposed to be  linearly independent and informationally complete (IC), we observe that  SQPT now is equivalent to deciding a unknown state
  with a set of product IC-POVM in  the $d^2$-dimensional Hilbert space.
   Following the general linear state tomography theory, we show that the product symmetric IC-POVM  should minimize the mean-square Hilbert-Schmidt distance between the estimator and the true states. So, an optimal SQPT can be realized by preparing both the inputs and the measurements as the symmetric IC-POVM.

keyword:~~~Standard quantum process tomography; The Cramer's rule; Linear state tomography.
\end{abstract}
 \maketitle

\section{introduction}

  The characterization of the evolution
of a quantum system is one of the main tasks to accomplish to
achieve quantum information processing. A general class of methods,
which have been developed in quantum information theory to
accomplish this task is known as quantum process tomography (QPT)-
for a review of quantum tomography, see Refs. [$1-3$].
The various protocols of getting the complete information about the quantum process  can be divided into two classes: The  standard quantum process tomography (SQPT) [1,4,5] which does not require any additionally quantum resources, and the so-called
ancilla-assisted quantum process tomography [6-8].

Usually, the complete characterization of $\chi$ matrix, which contains the complete information about a unknown quantum chnnel,  is a
non-scalable task: For the  $N$ $d$-level system, there are about
$d^{4N}$ elements to be decided. Recently, a series of works, where the ancilla system is also required, have
demonstrated that it is possible to extract partial but nevertheless
relevant information about the quantum process in an efficient and
scalable way [9-15]. These approaches share an essential feature
that: They are based on the idea that the tomography of a quantum
map can be efficiently performed by studying certain properties of a
twirling of such a map. Another method, the so-called direct
characterization of quantum dynamics (DCQD), was also constructed
for use in partial characterization  of quantum dynamics [16-17].
In experiment, the selective and efficient quantum process tomography has been recently carried out with
single photons [18].

With the aid of a clean ancilla system, the approach of ancilla-assisted quantum process tomography is to encode all information about the transformation into a single bipartite
system-ancilla quantum state, and thus completely reduce the problem to that of quantum state tomography. In general, there three known methods  of performing QST: The maximum likelihood [19-20], the   Bayesian mean method [22-25], and the linear state tomography [26-27].
Our present work is motivated by such a simple reasoning: To carry out  the
 ( complete and partial ) ancilla-assisted  QPT  in experiment,  one should at first perform a complete QPT
experiment to show that the ancilla system is (nearly) perfect. To
avoid the logical recycling, this complete QPT experiment should be
realized by the SQPT protocol which does not require any additional
resources. In this aspect,  the  SQPT is not only the unique method, in which the quantum process tomography can be directly performed, but it  also plays an important  role in realizing  the various kinds of ancilla-assisted SQPT in experiment.

In present work, we shall focus on the following two topics about the SQPT. At first, in the SQPT protocol for   a $d$-dimensional system, one should encounter a problem in solving  of a set of $d^4$ linear equations in order to get the matrix containing the complete information about the unknown quantum channel. Until now, the general form of the solution is unknown. And a long existed conviction is that the solutions are not unique. Here, we shall develop a self-consistent scheme,  in which bounded linear operators are presented by vectors, to construct the set of linear equations. With the famous Cramer's rule for the  set of linear equations, we are able to give the general form of the solution and show that it is unique.
In second, the central idea of the SQPT is to prepare a set of linearly independent inputs and measuring the outputs via the quantum state tomography (QST).   Letting  the inputs and the measurements be prepared by  two sets of the rank-one
   positive-operator-valued measures [POVMs], where each POVM  is supposed to be  linearly independent and informationally complete (IC), we observe that  SQPT now is equivalent to deciding a unknown state
  with a set of product IC-POVM in  the $d^2$-dimensional Hilbert space. This observation makes it possible for us to judge which choices of the
  inputs and measurements are  optimal for SQPT by
  following  the general discussion of Scott [26-27].

The content of present work can be divided into following parts. At
section 2,  we shall introduce the isomorphism where bounded operators are related to vectors. Several  super operators and some  relations
among them should be carefully defined and proven there. In section 3, we shall represent the SQPT theorem  in the vector denotation. The general solution of the set of $d^4$ linear equations should be given. Its uniqueness should be proven by the well-known Cramer's rule. In section 4, by jointing our SQPT protocol with the linear state tomography theorem, we show that the symmetric IC-POVM, which has been introduced  [28], should be used as the inputs and the measurements operators of SQPT. Under the linear quantum state tomography, the biproduct SIC-POVM not only feature a simple state reconstruction formula but also minimize mean-square Hilbert-Schmidt distance between the estimator and the true states.
In section 5, for the optimal case, we offer a simple formula to get the matrix containing the complete information of the quantum channel. Finally, we end our work with a short discussion.

\section{The vector denotation}

Consider a d-dimensional Hilbert space $H$ with $\{\vert i\rangle\}_{i=1}^d$
 the basis for it. We could define a set of operators $\{C_{ij}\}_{i,j=1}^d$,
 \begin{equation}
 C_{ij}=\vert i\rangle\langle j\vert, \mathrm{Tr} [C^{\dagger}_{ij}C_{kj}]=\delta_{ik}\delta_{jl}.
 \end{equation}
This set of operators formulate a orthogonal basis on the state space. Any $d\times d$ matrix in $H$, say $A$, can be expanded with it,
\begin{equation}
A=\sum_{i,j=1}^{d}A_{ij}C_{ij}, A_{ij}=\mathrm{Tr} [C_{ij}^{\dagger}A],
\end{equation}
with $A_{ij}$ the matrix elements of $A$. In the following argument, we will make use of the convenient isomorphism in which bounded linear operators are presented by vectors and denoted with the symbol $\vert\cdot\rangle\rangle$. It should be noted that the ways of defining $\vert \cdot\rangle\rangle $ are  not limited [29-33]. Although our scheme  is similar with the one introduced  in [33], for the self-consistency of present work, it will be presented  in an independent and complete form.

\emph{Theorem 1. } Let $\vert S_+\rangle\rangle$ be the maximally entangled state in $H^{\otimes2}$,
\[\vert S_{+}\rangle\rangle =\frac{1}{\sqrt{d}}\sum_{k=1}^d\vert k\rangle\otimes\vert k\rangle.\]
If an isomorphism between the $d\otimes d$ matrix $A$ and a $d^2$-dimensional
 vector $\vert A\rangle\rangle$   is defined as
 \begin{equation}
 \vert A\rangle\rangle =\sqrt{d} A\otimes \mathrm{I}_{d}\vert S_+\rangle\rangle,
  \end{equation}
 there should be
 \begin{equation}
 \vert A\rangle\rangle=\sum_{i,j=1}^d A_{ij}\vert i\rangle\otimes \vert j\rangle\equiv\sum_{i,j=1}^d A_{ij} \vert ij\rangle\rangle.
 \end{equation}
\emph{Proof. } From the definition of the isomorphism above, one may check that each  operator in (1) is related to the basis vector in
$H^{\otimes2}$,
\[\vert C_{ij}\rangle\rangle=\vert i\rangle\otimes\vert j\rangle\equiv \vert ij\rangle\rangle.\] Since that the isomorphism is linear, we shall get
the form in (4) where the matrix elements $A_{ij}$ are related to the  expanding coefficients of the vector $\vert A\rangle\rangle$.

\emph{Theorem 2.} Suppose that  $A$ , $B$, and $\rho$  are three arbitrary bounded matrices in ${H}$, there should be
\begin{eqnarray}
\mathrm{Tr}[A^{\dagger}B]=\langle\langle A\vert B\rangle\rangle,\\
\vert A\rho B\rangle\rangle =A\otimes B^{\mathrm{T}}\vert \rho\rangle\rangle.
\end{eqnarray}
\emph{Proof}. The equation (5) can be proved with the definition of the vector in (4),
\[\langle\langle A \vert B\rangle\rangle=(\sum_{i.j=1}^d A_{ij}\vert ij\rangle\rangle)^{\dagger}\sum_{k.l=1}^d B_{kl}\vert kl\rangle\rangle
=\sum A^{\dagger}_{ji}B_{ij}=\mathrm{Tr}[A^{\dagger}B].\]
The equation in (6) can be verified in the way like,
\begin{eqnarray}
A\otimes B^{\mathrm{T}}\vert \rho\rangle\rangle&=& (\sum_{i,k+1}^d A_{ik}\vert i\rangle\langle k\vert)\otimes( \sum_{j,l=1}^d
B_{lj}\vert j\rangle\langle l\vert)\sum_{m.n=1}^d\rho_{mn}\vert \rho\rangle\rangle \nonumber\\
&=&\sum_{i,j,k,l=1}^d A_{ik} \rho_{kl}B_{lj}\vert ij\rangle\rangle\nonumber\\
&=&\vert A\rho B\rangle\rangle.\nonumber
\end{eqnarray}

Let the quantum channel of the $H$ be described with a set of Kraus operators, $\varepsilon:\{E^{m}\}$. With $\rho$ for the input of the channel, the output $\varepsilon(\rho)$ should be decided by the Kraus operators,
\begin{equation}
\varepsilon(\rho)=\sum_{m}E^{m}\rho(E^{m})^{\dagger}, \sum_m(E^m)^{\dagger}E^m=\mathrm{I}_d.
\end{equation}
 The channel is supposed to be trace preserving.  For this set of Kraus operators,  a supper operator $\chi^{\mathrm{c}}$ is defined as,
\begin{equation}
\chi^{\mathrm{c}}=\sum_m\vert E^m\rangle\rangle\langle\langle E^m\vert\equiv d\cdot\varepsilon\otimes \mathrm{I}_d(\vert S_+\rangle\rangle\langle\langle S_+\vert),
\end{equation}
which is  Hermitian and positive. Its matrix elements, $\chi^{\mathrm{c}}_{ij;kl}=\langle\langle ij\vert \chi^{\mathrm{c}}\vert kl\rangle\rangle$, take the form
\begin{equation}
\chi^{\mathrm{c}}_{ij;kl}=\sum_m E^m_{ij}(E^m_{kl})^*.
\end{equation}
If these coefficients have been  known, the output in (7) should  be decided with the given input,
\begin{equation}
\varepsilon(\rho)=\sum_{ij;kl}\chi^{\mathrm{c}}_{ij;kl}\vert i\rangle \langle j\vert \rho\vert l\rangle\langle k\vert.
\end{equation}

Noting $\mathrm{Tr}[\chi^{\mathrm{c}}]=d$, we can also define a normalized density operator $\rho_{\varepsilon}$,
\begin{equation}
\rho_{\varepsilon}=\frac{1}{d}\chi^{\mathrm{c}}, \rho_{\varepsilon}=\varepsilon\otimes \mathrm{I}_d(\vert S_+\rangle \rangle\langle\langle S_+\vert).
\end{equation}
From it, we see that the tomography of $\chi^{\mathrm{c}}$ is equivalent  to determining the unknown state $\rho_{\varepsilon}$ in $H^{\otimes 2}$.
With $\{\vert ij\rangle \rangle\}_{i,j=1}^d$ the basis of $H^{\otimes 2}$,  the set of operators $\{\tilde{C}_{ij;kl}\}_{i,j,k,l=1}^d$, which are  defined as
\begin{equation}
\tilde{C}_{ij;kl}=\vert ij\rangle\rangle\langle \langle kl\vert\equiv\vert i\rangle \langle k\vert\otimes \vert j\rangle\langle l\vert,
\end{equation}
should formulate an orthogonal operator basis. Any arbitrary $d^2\otimes d^2$ matrix, say, $\Gamma$, can be expanded with it
\begin{equation}
\Gamma=\sum_{i,j,k,l=1}^d\Gamma_{ij;kl}\tilde{C}_{ij;kl}, \Gamma_{ij;kl}=\mathrm{Tr}[\tilde{C}_{ij;kl}^{\dagger}\Gamma],
\end{equation}
with $\Gamma_{ij;kl}$ the matrix element of $\Gamma$, $\Gamma_{ij;kl}=\langle \langle ij \vert \Gamma\vert kl \rangle\rangle$.  Jointing (8-12) together, we shall find the result
\[\sum_mE^m_{ij}(E^m_{kl})^*=d\cdot\mathrm{Tr}[\tilde{C}^{\dagger}_{ij;kl}\rho_{\varepsilon}]=d\cdot \mathrm{Tr}[\vert k\rangle\langle i\vert\otimes \vert l\rangle\langle j\vert\rho_{\varepsilon}]\]
and rewrite (10) with
\[\varepsilon(\rho)=d\sum_{ij;kl}\mathrm{Tr}[\vert k\rangle\langle i\vert\otimes \vert l\rangle\langle j\vert\rho_{\varepsilon}]\vert i\rangle \langle j\vert \rho\vert l\rangle\langle k\vert\]
 where the so-called \emph{Jamiolkowski isomorphism }is recovered here [34].

 As a direct application of (6), we find that the two vectors, $\vert  \varepsilon(\rho)\rangle\rangle$ and $\vert \rho\rangle\rangle $, are related by
 \[\vert \varepsilon(\rho)\rangle\rangle=\sum_{m}\vert E^m\rho (E^m)^{\dagger})=\sum E^m\otimes (E^m)^*\vert \rho\rangle\rangle.\]
 For convenience, we could introduce a super operator $\lambda^{\mathrm{c}}$,
 \begin{equation}
 \lambda^{\mathrm{c}}=\sum_m E^m\otimes (E^m)^*,
 \end{equation}
 and let the two vectors, which corresponds to the output and input of the quantum channel respectively, be related with a simple formula,
  \begin{equation}
  \vert \varepsilon(\rho)\rangle\rangle=\lambda^{c}\vert \rho\rangle\rangle.
  \end{equation}.

For the same set of Kraus operators $\{E^m\}$, we have introduced two different super operator, $\chi^{\mathrm{c}}$ in (8) and $\lambda^{\mathrm{c}}$ in (14), here. There is a question: In which way these two super operators are related? To answer
  this question,
we shall at fist define another isomorphism where a $d^2\otimes d^2$ matrix $\Gamma$ is related to the vector $\vert \Gamma)$ in $H^{\otimes 4}$.

\emph{ Theorem 3.} Let $\vert \tilde{S}_+)$ be a maximally entangled states in $H^{\otimes 4}$,
\[\vert \tilde{S}_+)=\frac{1}{d} \sum_{i,j=1}^d \vert ij\rangle\rangle\otimes \vert ij\rangle\rangle,\]
we could define a vector $\vert  \Gamma)$,
\begin{equation}
\vert \Gamma)=d \cdot\Gamma\otimes \mathrm{I}_{d^2}\vert\tilde{ S}_+).
\end{equation}
With $\Gamma_{ij;kl}$ the matrix elements of $\Gamma$, $\Gamma_{ij;kl}\equiv \langle\langle ij\vert \Gamma\vert kl\rangle\rangle$, there should be
\begin{equation}
\vert \Gamma)=\sum_{i,j,k,l=1}^d \Gamma_{ij;kl}\vert ij;kl), \vert ij;kl)\equiv \vert ij\rangle\rangle\otimes \vert kl\rangle\rangle.
\end{equation}
\emph{Proof.} It can be shown that each  basis operator $\tilde{C}_{ij;kl}$ in (12) should be related to the basis vector $\vert ij;kl)$,
\[\vert \tilde{C}_{ij;kl})=\vert ij;kl)\equiv\vert ij\rangle\rangle\otimes \vert kl\rangle\rangle.\]
With the known fact that the isomorphism is linear, we can get (17) by using the result in (13).

It should be noted that the isomorphism in (16) differs from  (4)  only in its dimension of the Hilbert apace. So, we may introduce the following
theorem without giving proof for it.

\emph{Theorem 4.} Suppose that $\Gamma$, $\Delta$, and $\Sigma$ are three arbitrary bounded matrices in $H^{\otimes 2}$, there should be
\begin{eqnarray}
\mathrm{Tr}[ \Gamma^{\dagger}\Delta]=(\Gamma\vert\Delta),\\
\vert \Gamma\Sigma\Delta)=\Gamma\otimes \Delta^{\mathrm{T}} \vert \Sigma ).
\end{eqnarray}

As the second step, we shall introduce  a super operator $\beta^{\mathrm{c}}$ in $H^{\otimes 4}$  with the following theorem:

\emph{Theorem 5. }
Let $\beta^{\mathrm{c}}$ be an operator in $H^{\otimes 4}$,
\begin{equation}
\beta^{\mathrm{c}}=\sum_{i,j,k,l=1}^d \vert ij;kl)(ik;jl\vert.
\end{equation}
It is Hermitian and unitary,
\begin{equation}
\beta^{\mathrm{c}}=(\beta^{\mathrm{c}})^{\dagger}=(\beta^{\mathrm{c}})^{-1}.
\end{equation}
The two vectors, $\vert \vert A\rangle\rangle\langle\langle B\vert )$ and $\vert A\otimes B^*)$, are related by
\begin{equation}
\beta^{\mathrm{c}}\vert \vert A\rangle\rangle\langle\langle B\vert )=\vert A\otimes B^*).
\end{equation}

\emph{Proof.} The property of $\beta^{\mathrm{c}}$ in (21) can be easily verified with the definition for it. Equation  (22)
can be proved as
\begin{eqnarray}
\beta^{\mathrm{c}}\vert \vert A\rangle\rangle\langle\langle B\vert )&=&\sum_{i,j,k,l=1}^dA_{ij}B^*_{kl}\beta^{\mathrm{c}}\vert \vert ij\rangle
\rangle\langle\langle kl\vert)\nonumber\\
&=&\sum_{i,j,k,l=1}^d A_{ij}B^*_{kl}\vert ik;jl)\nonumber\\
&=&\sum_{i,j,k,l}^d A_{ij}B^*_{kl}\vert \vert i\rangle\langle j\vert\otimes \vert k\rangle\langle l\vert)\nonumber\\
&=&\vert A\otimes B^*).\nonumber
\end{eqnarray}
  Finally, as  a direct conclusion based on  the  results above,  the answer of our question is known:  The two super operators, $\chi^{\mathrm{c}}$ and $\lambda^{\mathrm{c}}$, are simply related by the equations in below,
 \begin{equation}
 \vert  \chi^{\mathrm{c}})=\beta^{c}\vert \lambda^{\mathrm{c}}),  \vert \lambda^{\mathrm{c}})=\beta^{\mathrm{c}}  \vert \chi^{\mathrm{c}}).
 \end{equation}.

After having defined the  two super operators, $\chi^{\mathrm{c}}$ and $\lambda^{\mathrm{c}}$ above, for  describing  the quantum channel, we shall continue to
introduce the  operators  which are used for  the  measurement. Suppose that $\{ P_\mu\}_{\mu=1}^{d^2}$ is a set of \emph{linearly independent} operators
( which are not necessary to be orthogonal with each other).    According to the isomorphism in (4), we shall get a set of linearly independent vectors $\{\vert P_{\mu}\rangle\rangle\}_{\mu=1}^d$. Following the idea of Chefles [35,36], it's convenient for us to introduce the
\emph{reciprocal state} $\vert R_{\mu}\rangle\rangle$, which is defined as that lies in $H^{\otimes 2}$ and is orthogonal to all $\vert P_{\nu}\rangle\rangle$ for $\mu\neq\nu$,
\begin{equation}
\langle\langle R_\mu\vert P_\nu\rangle\rangle=\delta_{\mu\nu}.
\end{equation}
Such a set of reciprocal states $\{\vert R_{\mu}\rangle\rangle\}_{\mu=1}^{d^2}$ always exist if and only if the states in $\{\vert P_{\mu}\rangle \rangle\}_{\mu=1}^{d^2}$ are linearly independent [35,36]. Before proceeding with the further discussion about these operators, we shall at first make our denotation  into a more compact  form. In present work, we suppose that all the calculations in $H^{\otimes 2}$ are carried out with the basis $\{\vert ij\rangle\rangle\}_{i,j=1}^d$. For  simplicity, we may introduce
the denotation,
\begin{equation}
\vert\mu\rangle\rangle\equiv\vert ij\rangle\rangle, \mu=(d-1)i+j,
\end{equation}
where the single Greek letter $\mu$ takes values from $1$ to $ d^2$. Therefore, one may use $\Gamma_{\mu;\nu}$ for the matrix elements of $\Gamma$.
 In the calculation, $\Gamma_{\mu;\nu}=\langle\langle ij\vert \Gamma\vert kl\rangle\rangle$, the actual values of the indices, $i$, $j$, $k$, and $l$,
 should be decided by the rule in (25). Now, let's return back to the two set of states in (24) and begin at the following theorem.

\emph{Theorem 6.} For the set of linearly independent operators $\{P_{\mu}\}_{\mu=1}^{d^2}$, the super operator $P$ is defined by,
\begin{equation}
P=\sum_{\mu=1}^{d^2}\vert P_{\mu}\rangle\rangle\langle\langle \mu\vert, \det(P)\neq 0.
\end{equation}
It has the properties that
\begin{eqnarray}
P\vert \mu\rangle\rangle=\vert P_{\mu}\rangle\rangle,\\
\vert R_{\mu}\rangle\rangle=(P^{-1})^{\dagger}\vert \mu\rangle\rangle ,\\
\sum_{\mu=1}^{D^2}\vert R_{\mu}\rangle\rangle\langle\langle P_{\mu}\vert=\mathrm{I}_{d^2}
\end{eqnarray}
\emph{ Proof.} The first equation can be get with the definition of $P$. Since that $\det(P)\neq 0$, it's inverse $P^{-1}$ is unique decided by
$P^{-1}=\sum_{\mu=1}^{d^2}\vert \mu\rangle\rangle \langle\langle R_{\mu}\vert$. Based on it,  the equation (28) can be verified. The final
result can be also verified with  (27) and (28): $\sum_{\mu=1}^{D^2}\vert R_{\mu}\rangle\rangle\langle\langle P_{\mu}\vert=\sum_{\mu=1}^{d^2}
(P^{-1})^{\dagger}\vert\mu\rangle\langle\mu\vert P^{\dagger}=\mathrm{I}_{d^2}.$

In fact, the set of reciprocals states introduced above can be also defined as the  \emph{canonical dual frame} corresponding to $\{\vert P_{\mu}\}_{\mu=1}^{d^2}$.

\emph{Theorem 7.} For a known set of linearly independent vectors $\{\vert P_{\mu}\rangle \rangle\}_{\mu=1}^{d^2}$, a super operator $\mathcal{F}$
 is defined by
 \begin{equation}
 \mathcal{F}=\sum_{\mu=1}^{d^2}\vert P_{\mu}\rangle\rangle\langle\langle P_{\mu}\vert, \det{F}\neq 0.
 \end{equation}
 The canonical dual frame corresponding to $\{P_{\mu}\}_{\mu=1}^{d^2}$ can be defined as
  \begin{equation}
 \vert R_\mu\rangle\rangle=\mathcal{F}^{-1}\vert P_{\mu}\rangle\rangle.
\end{equation}
From these definitions, there should be $\langle\langle R_{\nu}\vert P_{\mu}\rangle\rangle=\delta_{\nu\mu}$.

\emph{ Proof.} Using the definition of the super operator $P$, one may get  $\mathcal{F}=PP^{\dagger}$. Since  that
 $\det{F}\neq 0$, the inverse of $F$ exists in an unique way, $F^{-1}=(P^{-1})^{\dagger}P^{-1}$. Jointed it with (27), there should be
 \[\mathcal{F}^{-1}\vert R_{\mu}\rangle\rangle=(P^{-1})^{\dagger}P^{-1}P\vert \mu\rangle\rangle=(P^{-1})^{\dagger}\vert\mu\rangle\rangle,\]
where the reciprocal states defined in (28) is recovered  here.

For the  special case where the set of operators $\{D_{\mu}\}_{\mu=1}^{d^2}$ are orthogonal with the algebra $ \mathrm{Tr}[D^{\dagger}_{\mu}D_{\nu}]=\delta_{\mu\nu}$, one may directly define a super operator $U$,
\begin{equation}
U=\sum_{\mu=1}^{d^2}\vert D_{\mu}\rangle \rangle\langle\langle\mu\vert.
\end{equation}
It can be easily verified that $U$ is unitary since that
\begin{equation}
U\vert \mu\rangle\rangle=\vert D_{\mu}\rangle\rangle
\end{equation}
in which it defines a mapping between the two set of basis vectors, $\{\vert \mu\rangle\rangle\}_{\mu=1}^{d^2}$ and
 $\{\vert D_{\mu}\rangle\rangle\}_{\mu=1}^{d^2}$, here.

\section{SQPT in the vector denotation}
In the first part of this section, we shall give a brief review of the SQPT theorem which is originally described in operator denotation. Then, via the isomorphism developed in the above section, we reexpress the SQPT in the vector denotation and show that $\chi$ matrix can be uniquely decided by the $\lambda$ matrix.

Suppose that $\{D_{\mu}\}_{\mu=1}^{d^2}$ is a set of fixed basis operators which satisfy the orthogonal condition $\mathrm{Tr}[D_{\mu}^{\dagger}D_{\nu}]=\delta_{\mu\nu}$. The operators  $E^m$ can be decomposed as
 $E^m=\sum_{\mu}e^m_{\mu}D_{\mu}$ with $e^m_{\mu}=\mathrm{Tr}[D^{\dagger}_{\mu}E^m]$ the expanding coefficients. Now, equation (7) can be rewritten as
 \begin{equation}
 \varepsilon(\rho)=\sum_{\mu,\nu=1}^{d^2}D_{\mu}\rho D^{\dagger}_{\nu}\chi_{\mu;\nu}.
 \end{equation}
 Where the coefficients $\chi_{\mu;\nu}$,
 \begin{equation}
 \chi_{\mu;\nu}=\sum_m\mathrm{Tr}[D^{\dagger}_{\mu}E^m](\mathrm{Tr}[D_{\nu}^{\dagger}E^m])^*,
 \end{equation}
 are the entries of a $ d^2\otimes d^2$ matrix $\chi$ which is Hermitian by definition.

 The central idea of SQPT is to prepare  $d^2$ linearly independent inputs $\{P_{\nu}\}_{\nu=1}^{d^2}$ and then measure the output
 $\varepsilon(P_{\nu})$. In addition, every   $\varepsilon(P_{\nu})$
can be expressed in terms of a linear combination of basis states,
\begin{equation}
 \varepsilon(P_{\nu})=\sum_{\mu=1}^{d^2}\lambda_{\mu;\nu}P_{\mu}.
 \end{equation}
Since $\varepsilon(P_{\nu})$ is known from the state tomography, $\lambda_{\mu;\nu}$ can be determined by standard linear algebraic algorithms. To
proceed, one may write
\begin{equation}
D_{\mu}P_{\delta}D_{\nu}^{\dagger}=\sum_{\gamma=1}^{d^2}\beta^{\mu;\nu}_{\gamma;\delta}P_{\gamma}
\end{equation}
where $\beta^{\mu;\nu}_{\gamma;\delta}$ are complex numbers which can be determined by standard algorithms from linear algebra given the $D_{\mu}$
and $P_{\gamma}$ operators. Combing the last two equations, we have
\begin{equation}
\sum_{\mu;\nu=1}^{d^2}\beta^{\mu;\nu}_{\gamma;\delta}\chi_{\mu;\nu}=\lambda_{\gamma;\delta}.
\end{equation}
One may take $\chi$ and $\lambda$ as vectors and $\beta$ as a $d^4\times d^4$ matrix with columns indexed by $\mu;\nu$ and rows by $\gamma;\delta$, and rewrite (38) as a set of linear equations
\begin{equation}
\beta\vec{\chi}=\vec{\lambda}.\end{equation}
The supper operator $\chi$ can be determined by inversion of this equation.

Until now, the general solution of $\chi$ has not been given.
And  it has been usually guessed that the solution is not unique. In the following argument, we shall show that the
vector dentation, which has been developed in above section,  offers a convenient tool in getting the solution of (39).

\emph{Theorem 8.} With $\chi^{\mathrm{c}}$ in (8) and the unitary $U$ in (32), a super operator $\chi$ is defined as
\begin{equation}
\chi=   U^{\dagger}\chi^{c}U,
\end{equation}
its matrix elements, $\langle\langle \mu\vert \chi\vert \nu\rangle\rangle$ , equal with the coefficients $\chi_{\mu;\nu}$ defined in (35).

\emph{Proof.} With Theorem 2, we may rewrite (35) as $\chi_{\mu;\nu}=\sum_m\langle\langle D_{\mu}\vert E^m \rangle\rangle\langle\langle  E_m\vert
D_{\nu}\rangle\rangle$. From the known definitions, $\chi^{\mathrm{c}}=\sum_{m}\vert E^m\rangle\rangle\langle\langle E^m\vert $ and
$\vert D_{\mu}\rangle\rangle=U\vert\mu\rangle\rangle$,  we shall get $\chi_{\mu;\nu}=\langle\langle \mu\vert \chi\vert \nu\rangle\rangle$.

 In the vector denotation, the  equation (36) can be rewritten
 $\vert \varepsilon(P_{\nu})\rangle\rangle=\sum_{\mu=1}^{d^2}\lambda_{\mu;\nu}\vert P_{\mu}\rangle\rangle$. With the known property of the reciprocal
states in (24), there should be $\lambda_{\mu;\nu}=\langle\langle R_{\mu}\vert \varepsilon(P_{\nu}\rangle\rangle$.
As it is shown in (15), $\vert \varepsilon(P_{\nu}\rangle\rangle=\lambda^{\mathrm{c}}\vert P_{\nu}\rangle\rangle$.
With the aid of theorem 7, we have $\langle\langle R_{\mu}\vert=\langle \langle \mu\vert P^{-1}$ and $\vert P_{\nu}\rangle\rangle=P\vert \nu\rangle\rangle$.
Jointing these results together,
we find (36) has an equivalent from
\begin{equation}
\lambda_{\mu;\nu}=\langle\langle \mu\vert P^{-1}\lambda^{\mathrm{c}}P\vert\nu\rangle\rangle.
\end{equation}
This result can expressed as the following theorem.

\emph{Theorem 9. } With the two super operators , $\lambda^{\mathrm{c}}$ in (14) and P in (26), a super operator $\lambda$ is defined
as
\begin{equation}
\lambda=P^{-1}\lambda^{\mathrm{c}}P.
\end{equation}
The coefficients $\lambda_{\mu;\nu}$ in (36) should be the matrix elements of it, $\lambda_{\mu;\nu}=\langle\langle\mu\vert \lambda\vert\nu\rangle\rangle$.

Note that  the  two super operators, $\chi$ and $\lambda$, are $d^2\times d^2$ matrices in $H^{\otimes 2}$. Via the isomorphism defined in theorem 3, we could introduce their corresponding vector, $\vert \chi)$ and $\vert \lambda)$ respectively,  in the enlarged Hilbert space $H^{\otimes 4}$. From (40), we could get the relation,  $\chi^{c}=U\chi U^{\dagger}$, and express it as
\begin{equation}
\vert \chi^{\mathrm{c}})=U\otimes U^*\vert \chi ).
\end{equation}
 In the similar way, we may rewrite (42) as
\begin{equation}
\vert \lambda)=P^{-1}\otimes P^{\mathrm{T}}\vert \lambda^{\mathrm{c}})
 \end{equation}
 Keeping in mind that $\vert \lambda^{c})$ is related to $\vert \chi^{\mathrm{c}})$ with the simple formula in (23), $\vert \lambda^{\mathrm{c})}=\beta^{\mathrm{c}} \vert \chi^{\mathrm{c}})$, we shall arrive at
 \begin{equation}
 \vert \lambda)=P^{-1}\otimes P^{\mathrm{T}}\beta^{\mathrm{c}}U\otimes U^*\vert \chi).
 \end{equation}
 Let's organize this result as the following theorem.

\emph{Theorem 10. } With the given super operators, $U$, $P$, and $\beta^{\mathrm{c}}$, a super operator $\beta$ in $H^{\otimes 4}$ can be defined
as
\begin{equation}
\beta=P^{-1}\otimes P^{\mathrm{T}}\beta^{\mathrm{c}}U\otimes U^*,
\end{equation}
 it relates $\vert \chi)$ to $\vert \lambda)$ in the way like
 \begin{equation}
 \vert \lambda)=\beta\vert \chi).
 \end{equation}
 The coefficients, $\beta^{\mu;\nu}_{\gamma;\delta}$ in (38), should be the matrix elements of $\beta$,
 $\beta^{\mu;\nu}_{\gamma;\delta}=(\gamma;\delta\vert \beta\vert \mu;\nu).$

\emph{ Proof.} At first, we could rewrite (37) in the vector denotation,
\[\vert D_{\mu}P_{\delta}D^{\dagger}_{\nu}\rangle\rangle=\sum_{\gamma=1}^{d^2}\beta^{\mu;\nu}_{\gamma;\delta}\vert P_{\gamma}\rangle\rangle.\]
With the definition of the reciprocal states in (24), there should be
\begin{equation}
\beta^{\mu;\nu}_{\gamma;\delta}=\langle\langle R_{\gamma}\vert D_{\mu}\otimes D_{\nu}^* \vert P_{\delta}\rangle\rangle.\nonumber
\end{equation}
We could show that the matrix elements of $\beta$ in (46) can be written in the way like
\begin{eqnarray}
(\gamma;\delta\vert \beta\vert \mu;\nu)&=&(\vert\gamma\rangle\rangle\langle\langle\delta\vert\vert
P^{-1}\otimes P^{\mathrm{T}}\beta^{\mathrm{c}}U\otimes U^{*}\vert\vert\mu\rangle\rangle\langle\langle \nu\vert)\nonumber\\
&=&((P^{-1})^{\dagger}\vert \gamma\rangle \rangle\langle\langle \delta\vert P^{\dagger}\vert \beta^{\mathrm{c}}\vert U\vert\mu\rangle\rangle\langle\langle\nu\vert U^{\dagger} )\nonumber\\
&=&(\vert R_{\gamma}\rangle\rangle\langle\langle P_{\delta}\vert\vert \beta^{\mathrm{c}}\vert \vert D_{\mu}\rangle\rangle\langle\langle D_{\nu}\vert)\nonumber\\
&=&(\vert R_{\gamma}\rangle\rangle\langle\langle P_{\delta}\vert\vert  D_{\mu}\otimes D_{\nu}^*)\nonumber\\
&=&\mathrm{Tr}[(\vert R_{\gamma}\rangle\rangle\langle\langle P_{\delta}\vert)^{\dagger}D_{\mu}\otimes D_{\nu}^*]\nonumber\\
&=&\langle\langle R_{\gamma}\vert D_{\mu}\otimes D_{\nu}^* \vert P_{\delta}\rangle\rangle.\nonumber
\end{eqnarray}
It should be noted that $\vert \mu;\nu)$ is equivalent to the vector $\vert ij;kl)$ with the constraints, as we have assumed in (25),
$\mu=(d-1)i+j$ and $\nu=(d-1)k+l$. Therefore, $\vert \mu\rangle\rangle\langle \langle \nu\vert\equiv\vert ij\rangle\rangle\langle\langle kl\vert
\equiv \tilde{C}_{ij;kl}$. In Theorem 3, it has been shown $\tilde{C}_{ij;kl}$ should be related to the basic vector $\vert ij; kl)$. Formally,
there is $\vert \vert \mu\rangle\rangle\langle\langle\nu\vert )=\vert \mu;\nu)$. Besides this result, theorem 4, theorem 7, and (33) have also been used in the derivation above.

The way of introducing (47) offers a well understanding about  the original formula in (39). It clearly shows why one may take $\chi$ and $\lambda$ as vectors and $\beta$ as a $d^4\times d^4$ matrix. At the end of this section, we shall prove that $\chi$ is uniquely determined if $\lambda$
has been given.

\emph{Theorem 11.} For the equation (47), $\vert \chi)$ is uniquely determined by $\vert \lambda)$,
\begin{equation}
\vert \chi)=\beta^{-1}\vert \lambda), \beta^{-1}=U^{\dagger}\otimes U^{\mathrm{T}}\beta^{\mathrm{c}}P\otimes (P^{-1})^{\mathrm{T}}.
\end{equation}

\emph{Proof.} The  equation (47) is nothing else but a system of $d^4$ linear equations. According to the famous
Cramer's rule [37]: \emph{If
$\texttt{\textbf{A}}\texttt{\textbf{x}}=\texttt{\textbf{b}}$ is a
system of n linear equations in n unknowns such that $\det
(A)\neq0$, then the system has a unique solution. }
So, if $\det(\beta)\neq 0$, we may conclude the solution of (47) is unique.
To derive the quantity  $\det(\beta)$, we shall use the following known results [37]: A. \emph{If~ $\Gamma$ and $\Delta$ are two
$d^2\times d^2$ matrices, there should be $\det(\Gamma\Delta)=\det(\Gamma)\det(\Delta)$};
B. \emph{The determinant of a $d\times d$ matrix $A$ equals the one of its transpose $A^{\mathrm{T}}$, $\det(A)=\det(A^{\mathrm{T}})$;}
 and
 C: \emph{If $A$ and $B$ are two $d\times d$ matrices, there should be $\det(A\otimes B)=\det(A)\det(B)$}.

Noting  that both $\beta^{\mathrm{c}}$ and $U$ are unitary matrices,
$\vert \det(\beta^{\mathrm{c}})\vert=\vert \det (U)\vert=\vert \det(U^{\dagger})\vert=1$, we shall have
\[\vert \det(\beta)\vert=\vert \det( P) \det(P^{-1})\vert.\]
Considering the fact that we have required the set of operators $\{P_{\mu}\}_{\mu=1}^{d^2}$ to be linearly independent,
both the determinants, $\det(P)$ and $\det(P^{-1})$,  are nonzero. Therefore, we conclude that the solution of (47) is unique since $\det(\beta)\neq 0$.
Finally, it should be mentioned that the isomorphism in Theorem 3 is a one-to-one mapping between the matrix and its corresponding vector.
So, it can be concluded  that $\chi$ is uniquely determined by $\lambda$.

\section{The optimal SQPT}

Since  that equation (48) can be divided into the following three equations
\begin{eqnarray}
\vert \chi )=U^{\dagger}\otimes U^{\mathrm{T}} \vert \chi^{\mathrm{c}}),\vert \chi^{\mathrm{c}})=\beta^{\mathrm{c}}\vert \lambda^{\mathrm{c}}),\\
\vert \lambda)=P^{-1}\otimes P^{\mathrm{T}}\vert \lambda^{\mathrm{c}}),
\end{eqnarray}
where $\chi$ and $\lambda$ matrices should be given is $\lambda^{\mathrm{c}}$ has been decided. In the following argument, we shall limit our attention to the task of deciding $\lambda^{\mathrm{c}}$ in experiment.

As we have shown, the central ideal of SQPT is to prepare $d^2$ linearly independent inputs $\{P_{\mu}\}_{\mu=1}^{d^2}$ and measure its output $\varepsilon(P_{\mu})$. Here, we shall focus on the situation where $\{P_{\mu}\}_{\mu=1}^{d^2}$ formulate a rank-one positive-operator-valued measure
(POVM),
\begin{equation}
P_{\mu}=w_{\mu}\vert\Phi_{\mu}\rangle\langle\Phi_{\mu}\vert, w_{\mu}=\mathrm{Tr}[P_{\mu}], \sum_{\mu=1}^{d^2}P_{\mu}=\mathrm{I}_d.
\end{equation}
while our discussion about it, which has presented   in the above argument, still holds here. The reason for this assumption is clear: The so defined $ P_{\mu}$ may be  easily prepared and its output $\varepsilon(P_{\mu})$ can be directly decided by the various  protocols of quantum states tomography (QST). It should be
noted that $\varepsilon(P_{\mu})$ has always been supposed to be known when our general solution in (48) is constructed. In other words, the SQPT  scheme itself does not offer  a  way of performing QST. Here, we suppose that the outputs, which lie in the $d$-dimensional Hilbert space,  are decided by the linear state tomography [26].

To describe the linear state tomography,  we should introduce another set of rank-one POVM $\{\bar{P}_{\mu}\}_{\mu=1}^{d^2}$,
\begin{equation}
\bar{P}_{\mu}=\bar{w}_{\mu}\vert\bar{\Phi}_{\mu}\rangle\langle\bar{\Phi}_{\mu}\vert, \bar{w}_{\mu}=\mathrm{Tr}[\bar{P}_{\mu}], \sum_{\mu=1}^{d^2}\bar{P}_{\mu}=\mathrm{I}_d.
\end{equation}
In special, we suppose that the operators in $\{\bar{P}_{\mu}\}_{\mu=1}^{d^2}$ are linearly independent. Introducing the super operator $\bar{P}$,
\begin{equation}
\bar{P}=\sum_{\mu}^{d^2}\vert \bar{P}_{\mu}\rangle\rangle\langle\langle \mu\vert, \bar{P}\vert \mu\rangle\rangle=\vert \bar{P}_{\mu}\rangle\rangle,
\end{equation}
The linearly independent condition holds iff $\det(\bar{P})\neq 0$.
We use $\vert \bar{R}_{\mu}\rangle\rangle)$ to denote the reciprocal states,
\begin{equation}
\vert \bar{R}_{\mu}\rangle\rangle=(\bar{P}^{-1})^{\dagger}\vert\mu\rangle\rangle,
\end{equation}
which should  have the property that
\begin{equation}
\sum_{\mu=1}^{d^2}\vert \bar{R}_{\mu}\rangle\rangle\langle\langle \bar{P}_{\mu}\vert=I_{d^2}.
\end{equation}
This property directly leads to the linear state reconstruction formula [26]
\begin{equation}
\vert \varepsilon(P_{\nu})\rangle\rangle=\sum_{\mu=1}^{d^2}\vert \bar{R}_{\mu}\rangle\rangle\langle\langle \bar{P}_{\mu}\vert \varepsilon(P_{\nu})\rangle\rangle
\end{equation}
or the equivalent form
\begin{equation}
\vert \varepsilon(P_{\nu})\rangle\rangle=\sum_{\mu=1}^{d^2}\omega_{\mu;\nu}\vert \vert \bar{R}_{\mu}\rangle\rangle,
\omega_{\mu;\nu}=\langle\langle \bar{P}_{\mu}\vert \varepsilon(P_{\nu})\rangle\rangle.
\end{equation}
Noting our task is to decide $\lambda^{\mathrm{c}}$ instead of a single output state. It can be realized by
preparing  a  series of  linear independent inputs $P_{\nu}$ and measuring each  output $ \varepsilon(P_{\nu})$ with the projective operator $\bar{P}_{\mu}$. There are altogether
$d^4$ such coefficients which can be collected as a $d^2\times d^2$ data matrix $\omega$,
\begin{equation}
\omega=\sum_{\mu,\nu=1}^{d^2}\omega_{\mu;\nu}\vert \mu\rangle\rangle\langle\langle\nu\vert.
\end{equation}
With our definitions of the super operators, $P$ in (27) and  $\bar{P}$ in (53), according to the relation (15), we could rewrite
the coefficient in (57) as
\begin{equation}
\omega_{\mu;\nu}=\langle\langle \mu\vert (\bar{P})^{\dagger}\lambda^{\mathrm{c}}P\vert\nu\rangle\rangle.
\end{equation}
Put it back into (58), there exists a simple formula connecting $\omega$ to $\lambda^{\mathrm{c}}$
\begin{equation}
\omega=\bar{P}^{\dagger}\lambda^{c}P
\end{equation}
which has an equivalent form
\begin{equation}
\lambda^{\mathrm{c}}=(\bar{P}^{-1})^{\dagger}\omega P^{-1}=\sum_{\mu;\nu=1}^{d^2}\omega_{\mu;\nu}(\bar{P}^{-1})^{\dagger}\vert\mu\rangle\rangle\langle\langle\nu\vert P^{-1}.
\end{equation}
Recalling our definitions of the reciprocal states, (28) and (54), we shall get one of main results of present work,
\begin{equation}
\lambda^{\mathrm{c}}=\sum_{\mu,\nu=1}^{d^2}\omega_{\mu;\nu}\vert \bar{R}_{\mu}\rangle\rangle \langle\langle P_{\nu}\vert,
\end{equation}
where the reciprocals states, $\vert \bar{R}_{\mu}\rangle\rangle$ and $\vert R_{\nu}\rangle\rangle$, are known if both the IC-POVMs,
$\{\bar{P}_{\mu}\}$ and  $\{P_{\nu}\}$,  have been designed at the beginning.  In the vector denotation, the above equation
takes the form
\begin{equation}
\vert \lambda^{\mathrm{c}})=\sum_{\mu,\nu=1}^{d^2}\omega_{\mu;\nu}\vert \vert\bar{R}_{\mu}\rangle\rangle \langle\langle P_{\nu}\vert).
\end{equation}
Applying the unitary transformation $\beta^{\mathrm{c}}$, which is defined in (21-22), on both sides of the above equation,   we shall get
\begin{equation}
\vert \chi^{\mathrm{c}})=\beta^{\mathrm{c}}\vert \lambda^{\mathrm{c}})=\sum_{\mu,nu=1}^{d^2}\omega_{\mu;\nu}\vert \bar{R}_{\mu}\otimes R_{\nu}^*).
\end{equation}
In the operator denotation, it should be
\begin{equation}
\chi^{\mathrm{c}}= \sum_{\mu,\nu=1}^{d^2}\omega_{\mu;\nu} \bar{R}_{\mu}\otimes R_{\nu}^*.
\end{equation}

Before showing which choices of the IC-POVMs are optimal in getting $\chi^{\mathrm{c}}$, we shall at first introduce some denotations to simplify our discussion. Formally, we define
\begin{equation}
\Pi_x=\bar{P}_{\mu}\otimes P^*_{\nu},
\end{equation}
 where the subindex $x$ takes value according to the rule
 \begin{equation}
 x=(d^2-1)\mu+\nu.
 \end{equation}
 One may easily check that$\{\Pi_x\}_{x=1}^{d^4}$  is a rank-one IC-POVM in $H^{\otimes2}$,
 \begin{equation}
 \sum_{x=1}^{d^2}\Pi_x=I_{d^2},
 \end{equation}
and the operators  in this set are linearly independent. At the same time, we define the operator $Q_x$,
\begin{equation}
Q_x=\bar{R}_{\mu}\otimes R_{\nu}^*.
\end{equation}

Note $(Q_x\vert \Pi_y)=\mathrm{Tr}[Q_x^{\dagger}\Pi_y]$, it can be easily verified that $\vert Q_x)$ should be the reciprocal states corresponding to $\vert \Pi_x)$. Therefore, there should be
\begin{equation}
 (Q_x\vert\Pi_y)=\delta_{xy}, \sum_{x=1}^{d^4}\vert Q_x)(\Pi_x\vert=\mathrm{I}_{d^4}.
\end{equation}
 With theses denotations in hand, one may simplify (64) into a more compact form,
 \begin{equation}
 \vert \chi^{\mathrm{c}})=\sum_{x=1}^{d^4} \omega (x)\vert Q_x)
 \end{equation}
 where $\omega (x)= \omega_{\mu;\nu}$ holds under the constraint in (67). Timing $(\Pi_y\vert $ on both sides of the above relation, one may
 get
 \begin{equation}
 \omega(y)=(\Pi_y\vert \chi^{\mathrm{c}}).
\end{equation}
  Noting $\mathrm{Tr}[\chi^c]=d$, we may introduce $\rho_{\varepsilon}$, which is defined in (11), for its normalized state. We rewrite the above two equations into such a form
  \begin{equation}
  \vert \rho_{\varepsilon})=\sum_{x=1}^{d^4}p(x)\vert Q_x), p(x)=(\Pi_x\vert \rho_{\varepsilon})\equiv\frac{1}{d}\omega_{\mu;\nu}.
  \end{equation}
  It's a standard form of linear state  reconstruction formula [27]: The unknown state $\rho_{\varepsilon}$ is measured with the rank-one IC-POVM $\{\Pi_x\}_{x=1}^{d^4}$, which lies in in the $H^{\otimes 2}$ Hilbert space, and reconstructed with the canonical dual frame $\{ Q_x\}_{x=1}^{d^4}$. From it, we are able to show which choice of
  $\Pi_x$ should be optimal by following the general discussion of Scott in [27,28]. Suppose that $ y_1, y_2,..., y_N$ are the outcomes of measurements on $N$ identical copies of the state $\rho_{\varepsilon}$. The estimate for the outcome probabilities is
  \begin{equation}
  \hat{\rho}_{\varepsilon}(x)=\hat{p}(x;y_1,y_2,...,y_N):=\frac{1}{N}\sum_{k=1}^N\delta(x,y_k),
  \end{equation}
 which gives a linear tomography estimate of $\rho_{\varepsilon}$ which obeys the expectation $E[\hat{p}(x)]=p(x)$.
 An elementary calculations shows hat the expected covariance for $N$ samples is
 \begin{equation}
E[(p(x)-\hat{p}(x)(p(x)-\hat{p}(x))]=\frac{1}{N}(p(x)\delta(x,y)-p(x)p(y)).\nonumber
\end{equation}
Now suppose that p(x) are outcomes probabilities for the IC-POVM $\{\Pi_{x}\}_{x=1}^{d^4}$   in (68). That is , $p(x)=(\Pi_x\vert \rho_{\varepsilon})$. The error in our estimate of $\rho_{\varepsilon}$,
\begin{equation}
\hat{\rho}_{\varepsilon}=\hat{\rho}_{\varepsilon}(y_1,y_2,...,y_N)\equiv\sum_{x=1}^{d^4}\hat{p}(x;y_1,...,y_N)Q(x).\nonumber
\end{equation}
as measured by the squared Hilbert-Schmidt distance, is
\begin{equation}
\vert\vert \rho_{\varepsilon}-\hat{\rho}_{\varepsilon}\vert\vert^2=\sum_{x,y=1}^{d^4}(p(x)-\hat{p}(x))(p(x)-\hat{p}(x))(Q_x\vert Q_y),\nonumber
\end{equation}
which has the expectation
\begin{eqnarray}
E[\vert\vert \rho_{\varepsilon}-\hat{\rho}_{\varepsilon}\vert\vert^2]&=&\frac{1}{N}((p(x)\delta(x,y)-p(x)p(y)))(Q_x\vert Q_y)\nonumber\\
&=&\frac{1}{N}\{(\sum_{x=1}^{d^4}p(x)(Q_x\vert Q_x)-\mathrm{Tr}[\rho_{\varepsilon}^2]\}\nonumber \\
&=&\frac{1}{N}(\Delta_p(Q)-\mathrm{Tr}[\rho_{\varepsilon}^2])\nonumber
\end{eqnarray}
where the quantity $\Delta_p(Q)$ is defined as
\begin{equation}
\Delta_p(Q)=\sum_{x=1}^{d^4}p(x)(Q_x\vert Q_x)\nonumber
\end{equation}
Since that $\rho_{\varepsilon}$ is unknown, it is the quantity $\Delta_p(Q)$ is now of interest. In general, it depends on the detail of $\rho_{\varepsilon}$. This dependence can be moved away if we set $\rho_{\varepsilon}=\rho_{\varepsilon}(\sigma,u)=u\sigma u^{\dagger}(Tr[\sigma]=1)$
and take (Haar) average over all $u\in u(d^2)$,
\begin{equation}
\Delta^{\mathrm{avg}}_p=\int_{u(d^2)}d\mu_{\mathrm{H}}(u)\sum_{x=1}^{d^4}\mathrm{Tr}[\Pi_x^{\dagger}u\sigma u^{\dagger}](Q_x\vert Q_x)\nonumber
\end{equation}
where $\mu_{\mathrm{H}}$ is the unit Haar measure. Using Shur's lemma,
\begin{equation}
\int_{u(d^2)}d\mu_{\mathrm{H}}(u)u\sigma u^{\dagger}=\frac{1}{d^2}\mathrm{I}_{d^2}.\nonumber
\end{equation}
we shall get the averaged quantity
\begin{equation}
\Delta^{\mathrm{avg}}_p=\frac{1}{d^2}\sum_{x=1}^{d^4}\mathrm{Tr}[\Pi_x](Q_x\vert Q_x).
\end{equation}
Let's express this formula in terms of the IC-POVMs, $\{P_{\mu}\}$ and $\{\bar{P_\nu}\}$, and their corresponding reciprocal states. For the IC-POVM $\{\bar{P}_{\mu}\}_{\mu=1}^{d^2}$, we define a super operator
\begin{equation}
\mathcal {\bar{K}}=\sum_{\mu=1}^{d^2}\frac{\vert \bar{P}_{\mu}\rangle\rangle\langle \bar{P}_{\mu}\vert}{\mathrm{Tr}[\bar{P}_{\mu}]}.\nonumber
\end{equation}
Its inverse should be
\begin{equation}
(\mathcal{\bar{K}})^{-1}=\sum_{\mu=1}^{d^2}\mathrm{Tr}[\bar{P}_{\mu}]\vert \bar{R}_\mu\rangle \rangle\langle\langle\bar{R}_{\mu}\vert\nonumber
\end{equation}
According to the lemma 17 in [26], there should be
\begin{equation}
\mathrm{Tr}[(\mathcal{\bar{K}})^{-1}]\geq d(d(d+1)-1)
\end{equation}
with equality \emph{iff $\{\bar{P}_{\mu}\}_{\mu=1}^{d^2}$ is a tight rank-one IC-POVM.}
The same argument is also suitable for the IC-POVM $\{ P^*_{\mu}\}_{\mu=1}^{d^2}$ ( iff  $\{ P_{\mu}\}_{\mu=1}^{d^2}$
is an IC-POVM ).  Introducing the super operator
\begin{equation}
\mathcal {K^*}=\sum_{\mu=1}^{d^2}\frac{\vert P^*_{\mu}\rangle\rangle\langle\langle P^*_{\mu}\vert}{\mathrm{Tr}[P^*_{\mu}]}\nonumber
\end{equation}
with its inverse to be
\begin{equation}
(\mathcal{{K}^*})^{-1}=\sum_{\mu=1}^{d^2}\mathrm{Tr}[{P}^*_{\mu}]\vert {R}^*_\mu\rangle \rangle\langle\langle{R}^*_{\mu}\vert.\nonumber
\end{equation}
Via the similar argument above, there should be
 \begin{equation}
\mathrm{Tr}[(\mathcal{{K^*}})^{-1}]\geq d(d(d+1)-1).\nonumber
\end{equation}
For the result in (75), one may easily find that
\begin{equation}
\Delta^{\mathrm{avg}}_p=\frac{1}{d^2}\mathrm{Tr}[(\mathcal{\bar{K}})^{-1}]\mathrm{Tr}[(\mathcal{{K^*}})^{-1}]\nonumber
\end{equation}
and the optimal value of it should be
\begin{equation}
\Delta^{\mathrm{avg}}_p=(d(d+1)-1)^2
\end{equation}
which can be achieved at if the both the IC-POVMs, $\{\bar{P}_{\mu}\}_{\mu=1}^{d^2}$ and $\{{P}^*_{\mu}\}_{\mu=1}^{d^2}$, have been chosen to be the
tight rank-one IC-POVM.

Let's organize the content of the present section in following way: To perform the SQPT for a $d-$dimensional system, we prepare a set of
$d^2$ linearly independent inputs   $\{{P}_{\nu}\}_{\nu=1}^{d^2}$, which are supposed to formulate an IC-POVM, and measure each
output $ \varepsilon(P_{\nu})$ with  anther set of IC-POVM $\{\bar{P}_{\mu}\}_{\mu=1}^{d^2}$.  All the $d^4$ experiment data are presented with the
matrix $\omega$ defined in (58). Within our present scheme, in which the SQPT theory is represented in the  vector denotation, we could give the
way of reconstructing $\lambda^{\mathrm{c}}$ in (62) and the one for $\chi^{c}$ in (64). As an interesting observation, we find that equation (64),
where $\vert \chi^{\mathrm{c}})$ can be viewed as  an unknown states in $H^{\otimes 4}$, is  just a special case of the linear state reconstruction  formula discussed by Scott in [27]. This observation makes it possible for us to get the optimal IC-POVMs, $\{{P}_{\nu}\}_{\nu=1}^{d^2}$ and $\{\bar{P}_{\mu}\}_{\mu=1}^{d^2}$, by following the general linear state tomography theory. The conclusion, in which  $\{\bar{P}_{\mu}\}_{\mu=1}^{d^2}$ should be a tight rank-one IC-POVM, can be viewed as an known result in [26]. In addition to it, we show that the conjugate of  inputs $\{{P}_{\nu}\}_{\nu=1}^{d^2}$ should also be prepared as a tight rank-one IC-POVM. This two requirements should make the statical error in reconstructing $\chi^{\mathrm{c}}$ to be minimal.

\section{The symmetric informationally complete POVM}

Under the condition of linearly independence, the tight rank-one IC-POVM  is unique [26]: It's the so called symmetric informationally  complete
POVM (SIC-POVM) introduced in [28]
\begin{equation}
P_{\mu}=\frac{1}{d}\vert\Phi_{\mu}\rangle\langle\Phi_{\mu}\vert\equiv\frac{1}{d}\Phi_{\mu}, \sum_{\mu=1}^{d^2}P_{\mu}=\mathrm{I}_{d}
\end{equation}
Where the normalized states $\vert \Phi_{\mu}\rangle$ have the following property [28]
\begin{equation}
\vert\langle \Phi_{\mu}\vert\Phi_{\nu}\rangle\vert^2=\frac{1+\delta_{\mu\nu}}{1+d}.
\end{equation}
From this definition, we see that $\{P^*_{\mu}\}_{\mu=1}^{d^2}$ should also be a SIC-POVM. In the following argument, we will suppose that
input and measurement IC-POVMs are the same one,
\[\bar{P}_{\mu}=P_{\mu}\].

The SIC-POVM defined above should also be the 2-design [26]
\begin{equation}
\sum_{\mu=1}^{d^2}P_{\mu}\otimes P_{\mu}=\frac{1}{d(d+1)}\sum_{j,k=1}^d \vert j\rangle\langle j\vert\otimes \vert k\rangle\langle k\vert+\vert j\rangle\langle k\vert\otimes \vert k\rangle\langle j\vert . \nonumber
\end{equation}
Let's take it as our starting point here to derive the reciprocal (canonical dual frame) operators $R_{\mu}$ needed in our reconstructing formula (65). Noting $P_{\mu}$ is a projective operator which has the property that $P^*=P^{\mathrm{T}}$. Therefore, performing the partial transposition
on above equation, we shall get
\begin{equation}
\sum_{\mu=1}^{d^2}P_{\mu}\otimes P^*_{\mu}=\frac{1}{d(d+1)}\sum_{j,k=1}^d \vert j\rangle\langle j\vert\otimes \vert k\rangle\langle k\vert+\vert j\rangle\langle k\vert\otimes \vert j\rangle\langle k\vert .
\end{equation}
In the vector denotation, it can be rewritten as
\begin{equation}
\sum_{\mu=1}^{d^2}\vert P_{\mu}\otimes P^*_{\mu})=\frac{1}{d(d+1)}\sum_{j,k=1}^d \vert jk;jk)+\vert jj;kk) \nonumber
\end{equation}
in which we have use our denotation in (17), $\vert ij;kl)=\vert ij\rangle\rangle\otimes \vert kl\rangle\rangle$. Performing the unitary transformation, $\beta^{c}$ in (20), on both sides of it, we shall get
\begin{equation}
\sum_{\mu=1}^ {d^2}\vert \vert P_{\mu}\rangle\rangle\langle\langle  P_{\mu}\vert)=\frac{1}{d(d+1)}\sum_{j,k=1}^d
\vert jk;jk)+\vert jj;kk) \nonumber
\end{equation}
Noting in  our definition of the isomorphism in theorem 3, the basis vector $\vert ij;kl)$ has a one-to-one mapping with the operator
$\tilde{C}_{ij;kl}$ in (12). Recalling our definition of the supper operator $\mathcal{F}$ in (30),
we shall get the formula
   \begin{equation}
\mathcal{F}=\sum_{\mu=1}^ {d^2} \vert P_{\mu}\rangle\rangle\langle\langle P_{\mu}\vert=\frac{1}{d(d+1)}\sum_{j,k=1}^d
\vert j\rangle \langle j\vert\otimes \vert k\rangle\langle k\vert +\vert j\rangle \langle k\vert\otimes \vert j\rangle\langle k\vert . \nonumber
\end{equation}
Using the relations that $\sum_{j,k=1}^d\vert j\rangle \langle j\vert\otimes \vert k\rangle\langle k\vert =\mathrm{I}_{d^2}$
and $\sum_{j,k=1}^d\vert j\rangle \langle k\vert\otimes \vert j\rangle\langle k\vert=\vert \mathrm{I}_d\rangle\rangle\langle \langle \mathrm{I}_d\vert$, we shall get
\begin{equation}
\mathcal{F}=\frac{1}{d(d+1)}(\mathrm{I}_{d^2}+\vert\mathrm{ I}_d\rangle\rangle\langle\langle \mathrm{I}_d\vert)
\end{equation}
with its inverse to be
\begin{equation}
\mathcal{F}^{-1}=d(d+1)\mathrm{I}_{d^2}-d\vert \mathrm{I}_d\rangle\rangle\langle\langle \mathrm{I}_d\vert.
\end{equation}
With equation (31), we find the  reciprocal states $\vert R_{\mu}\rangle\rangle$,
\begin{equation}
\vert R_{\mu}\rangle\rangle=\mathcal{F}^{-1}\vert P_{\mu}\rangle\rangle=d(d+1)\vert P_{\mu}\rangle\rangle-d\mathrm{Tr}[P_{\mu}]\vert \mathrm{I}_d\rangle\rangle\nonumber
\end{equation}
From our definition of $P_{\mu}$ in (65), we shall get a compact form of $R_{\mu}$,
\begin{equation}
R_{\mu}=(d+1)\Phi_{\mu}-\mathrm{I}_{d},
\end{equation}
which can be used to reconstruct $\chi^{\mathrm{c}}$ in (65). It's meaningful to calculate another super operator
\begin{equation}
\mathcal{K}\equiv\sum_{\mu=1}^{d^2}\frac{\vert P_{\mu}\rangle\rangle\langle\langle P_{\mu}\vert}{Tr[P_{\mu}]}=\frac{1}{(d+1)}(\mathrm{I}_{d^2}+\vert\mathrm{ I}_d\rangle\rangle\langle\langle \mathrm{I}_d\vert)
\end{equation}
with it's inverse to be
\begin{equation}
\mathcal{K}^{-1}=(d+1)\mathrm{I}_{d^2}-\vert \mathrm{I}_d\rangle\rangle\langle\langle \mathrm{I}_d\vert.
\end{equation}
Formally, it can be rewritten as the  mixture  of a set of basis vectors $\{\vert e_{\mu}\rangle\rangle\}_{\mu=1}^{d^2}$, which satisfy the constraints that $\vert e_1\rangle\rangle= \frac{1}{\sqrt{d}}\sum_{j=1}^d\vert jj\rangle\rangle$ and $\langle\langle e_{\mu}\vert e_{\nu}\rangle\rangle=\delta_{\mu\nu}$,
\begin{equation}
\mathcal{K}^{-1}=\vert e_1\rangle\rangle\langle\langle e_1\vert+(d+1)\sum_{\mu=2}^{d^2}\vert e_{\mu}\rangle\rangle\langle\langle e_{\mu}\vert.
\end{equation}
Obviously, its eigenvalues  should $\lambda_1=1$ and $\lambda_{\mu}=d+1$ for $\mu\neq 1$ . The trace of it should be
\begin{equation}
\mathrm{Tr}[\mathcal{K}^{-1}]=1+(d+1)(d^2-1)=d(d(d+1)-1)
\end{equation}
in which the lower bound of (76) is achieved here. By the way, one may check that $\mathcal{K}^*=\mathcal{K}$, the above bound also holds for the
SIC-POVM $\{P_{\mu}^*\}$.

With the assumption that both the inputs and measurement operators are the SIC-POVM defined in (78), we may rewrite the coefficients $\omega_{\mu;\nu}$ in (57) as
\begin{equation}
\omega_{\mu;\nu}=\frac{1}{d^2}\mathrm{Tr}[\Phi_{\mu}^{\dagger}\varepsilon(\Phi_{\nu})].
\end{equation}
Putting (83) back into (65), we shall get our reconstruction formula for $\chi^{\mathrm{c}}$,
\begin{equation}
\chi^{\mathrm{c}}=-\mathrm{I}_{d^2}+\sum_{\mu,\nu=1}^{d^2}\omega_{\mu;\nu}[(d+1)^2\Phi_{\mu}\otimes \Phi_{\nu}^*-(d+1)\Phi_{\mu}\otimes \mathrm{I}_{d}],
\end{equation}
in which we have used the relations $\sum_{\mu;\nu=1}^{d^2}\omega_{\mu;\nu}=d$ and $\sum_{\mu=1}^{d^2}\omega_{\mu;\nu}=1/d$. If the quantum channel is \emph{unital}, where $\varepsilon(\mathrm{I}_d)=\mathrm{I}_d$, the above formula can be simplified as
\begin{equation}
\chi^{\mathrm{c}}=-(d+2)\mathrm{I}_{d^2}+(d+1)^2\sum_{\mu,\nu=1}^{d^2}\omega_{\mu;\nu}\Phi_{\mu}\otimes \Phi_{\nu}^*.
\end{equation}

In the end of this section, we shall offer an example to show how our reconstruction formula works. Let's consider the \emph{depolarizing channel},
where an arbitrary state $\Phi$ should have its output to be
\begin{equation}
\varepsilon(\Phi)=(1-q)\Phi+\frac{q}{d}\mathrm{I}_{d},
\end{equation}
with $q$ a positive parameter taking values form 0 to 1. One may check it's an unital quantum channel since $\varepsilon(\mathrm{I}_d)=\mathrm{I}_d$.
Via a simple calculation, we find
\begin{equation}
\mathrm{Tr}[\Phi^{\dagger}_{\mu}\varepsilon(\Phi_{\nu})]=\frac{d(1-q)}{d+1}\delta_{\mu\nu}+\frac{d+q}{d(d+1)}.\nonumber
\end{equation}
Jointing it with equations (80) and (90), we find that $\chi^{c}$ is now a Werner-type state,
\begin{equation}
\chi^{c}=d\rho_{\mathrm{Werner}}, \rho_{\mathrm{Werner}}=(1-q)\vert S_+\rangle \rangle \langle\langle S_+\vert +\frac{q}{d^2}\mathrm{I}_{d^2},
\end{equation}
where $\vert S_+\rangle\rangle $ is the maximally entangled state defined in Theorem 1.
\section{discussion}
In the ancilla-assisted QPT, one may prepare a maximally entangled state $ \vert S_+\rangle\rangle=\frac{1}{\sqrt{d}}\sum_{j=1}^d\vert j\rangle\vert j\rangle$
as the input with its corresponding output  defined as
\begin{equation}
 \rho_{\varepsilon}=\varepsilon\otimes \mathrm{I}_d(\vert S_+\rangle \rangle\langle\langle S_+\vert).\nonumber
\end{equation}
If this output has been decided  with the quantum state tomography, the information of the quantum process, which is now described by the
the matrix $\chi^{\mathrm{c}}=\sum_{m}\vert E^{m}\rangle\rangle\langle\langle E^{m}\vert$, should also be known as
 $\chi^{\mathrm{c}}=d\rho_{\varepsilon}$.

As we have shown,  our  present protocol is equivalent to deciding $\rho_{\varepsilon}$ with a product SIC-POVM. In  [27], Scott has demonstrated that the so-called \emph{unitary 2-designs}  are the optimal choice for the ancilla-assisted QPT in getting $\rho_{\varepsilon}$. For the expectation
\[E[\vert\vert \rho_{\varepsilon}-\hat{\rho}_{\varepsilon}\vert\vert^2]=\frac{1}{N}(\Delta_p(Q)-\mathrm{Tr}[\rho_{\varepsilon}]^2),\]
the unitary 2-designs should offer a optimal value
\begin{equation}
\Delta^{\mathrm{opt}}_p(Q)=d^{4}-d^{2}+\frac{1}{d^2},
\end{equation}
for the general quantum channel ( an equivalent  result can be also found in [38]).
There is a simple inequality between this optimal value and our result $\Delta^{\mathrm{avg}}_p$ in (77):
\begin{equation}
\frac{\Delta^{\mathrm{opt}}_p}{\Delta^{\mathrm{avg}}_p}>\frac{d-1}{d+1}.
\end{equation}
When $d$, the dimension  of the system, is increasing, the ratio above  should approach unit. A related  discussion about the product SIC-POVM can be also found  in [39], where the authors shows that there is only a marginal efficiency advantage of the joint SIC-POVM over the product IC-POVM in  the state tomography.

In the present work, we have assumed that the outputs are measured with a set of linear independent operators. This assumption make it possible for us to introduce the  Chefles' reciprocal states which is convenient for our calculation. However,
a set of nonlinear independent IC-POVMs can be also used in the quantum state tomography. How to generalize our  scheme for such situations, should be studied  in future.

Finally, let's end our work with a short conclusion.
 At first,  we  developed a self-consistent scheme,  in which bounded linear operators are presented by vectors, for constructing  the set of linear equations in order to get the matrix containing the complete information about the quantum process.  We  proved that the solution of this set of linear equations  is unique.
In second, letting  the inputs and the measurements be prepared by  two sets of the rank-one
   positive-operator-valued measures [POVMs], where each POVM  is supposed to be  linearly independent and informationally complete (IC), we observe that  SQPT now is equivalent to deciding a unknown state
  with a set of product IC-POVM in  the $d^2$-dimensional Hilbert space. Following the general linear state tomography theory, we show that the product symmetric IC-POVM  should minimize the mean-square Hilbert-Schmidt distance between the estimator and the true states. So, an optimal SQPT can be realized by preparing both the inputs and the measurements as the symmetric IC-POVM.

\end{document}